\documentstyle[12pt]{article}
\catcode`\@=11 \@addtoreset{equation}{section}\catcode`\@=12

\newcommand{\be}{\begin{equation}}
\newcommand{\ee}{\end{equation}}
\newcommand{\ba}{\begin{eqnarray}}
\newcommand{\ea}{\end{eqnarray}}

\newcommand{\lra}{\longrightarrow}

\newcommand{\ra}{\rightarrow}

\newcommand{\p}{\partial}

\newcommand{\btd}{\bigtriangledown}
\newcommand{\btu}{\bigtriangleup}

\begin{document}
\thispagestyle{empty}

\begin{center}

               RUSSIAN GRAVITATIONAL ASSOCIATION\\
               CENTER FOR  GRAVITATION  AND FUNDAMENTAL METROLOGY\\
               ALL-RUSSIAN RESEARCH INSTITUTE OF METROLOGICAL
               SERVICE\\

\end{center}
\vskip 4ex
\begin{flushright}                              RGA-VNIIMS-004/95\\
                                                gr-qc/9507056
\end{flushright}
\vskip 15mm

\begin{center}
{\large\bf
On singular solutions in multidimensional gravity}

\vskip 5mm
{\bf
Vladimir D. Ivashchuk and Vitaly N. Melnikov }\\
\vskip 5mm
     {\em Center for Gravitation and Fundamental Metrology, VNIIMS\\
     3/1 M.Ulyanovoy str., Moscow, 117313, Russia}\\
     e-mail: ivas@cvsi.uucp.free.net \\
\end{center}
\vskip 10mm

ABSTRACT

It is proved that the Riemann tensor squared is divergent as
$\tau  \ra 0$  for a wide class of cosmological metrics with
non-exceptional Kasner-like behaviour of scale factors as
$\tau  \ra 0$, where $\tau$ is synchronous time.
Using this result it is shown that non-trivial generalization
of the spherically-symmetric Tangherlini solution to the case of
$n$ Ricci-flat internal spaces \cite{FIM}  has a divergent Riemann tensor
squared as  $R  \ra R_0$, where $R_0$ is parameter of
length of the solution. Multitemporal naked singularities are also
considered.

\vskip 10mm


\vskip 30mm

\centerline{Moscow 1995}
\pagebreak

\setcounter{page}{1}

\pagebreak

\section{Introduction}

In the recent decade there has been a great interest to multidimensional
models of classical and quantum gravity (see, for example,
\cite{Vl1}-\cite{M} and references therein). This interest was stimulated
mainly by the investigations on the theory of supergravity and
superstrings \cite{GSW}, i.e.  the ideas of the unification of
interactions gave a "new life" to the ideas of Kaluza and Klein.

As it was shown in papers \cite{IM1}-\cite{IM3} devoted to
multidimensional cosmological models with perfect fluid a large variety of
the exact solutions have a Kasner-like asymptotical behaviour for small
values of the synchronous time parameter  $\tau  \ra 0$. Here we do not
consider the exceptional solutions such as exponential and power-law
inflationary solutions \cite{IM3}. The solutions with oscillatory (or
stochastic behaviour) as $\tau  \ra 0$ also are not covered by the
presented scheme.

In this paper we use the Riemann tensor squared as an indicator of the
singular behaviour of the cosmological solutions as $\tau  \ra 0$.
In Sec. 2 we present an explicit formula for the quadratic invariant
(Riemann tensor squared)
of the metric defined on the product of spaces with the scale
factors depending on the points of the first space. In Sec. 3
the  quadratic invariant is presented for a wide class of cosmological
metrics describing the evolution of $n$ spaces of arbitrary dimensions.
It is proved (see Proposition 2) that, when  all spaces are 1-dimensional,
the Riemann tensor squared is positive and divergent as $t = \tau  \ra 0$
for all non-trivial (non-Milne-like) configurations. In Subsec. 3. 2 this
result is generalized to the case of Ricci-flat spaces.  In Subsec. 3.3
the main theorem concerning the divergency of the Riemann tensor squared
for a wide class of cosmological metrics with non-exceptional Kasner-like
behaviour of scale factors as $\tau  \ra 0$ is proved. In Sec. 4 we apply
the obtained result to the generalization of the spherically-symmetric
Tangherlini solution \cite{T} to the case of $n$ Ricci-flat internal
spaces \cite{FIM}.  In Subsec. 4.1 the  multitemporal generalization of
the Tangherlini solution is considered \cite{IM4}.  This solution is shown
to describe the (multitemporal) naked singularity when the parameters are
non-exceptional.

\section{General formalism}

Let $(M,g)$ be a manifold $M$  with the metric $g$. The squared Riemann
tensor
\be  %2.1
{I}[g] \equiv {R_{MNPQ}}[g] {R^{MNPQ}}[g]
\ee
is a smooth real-valued function on $M$. For smooth function
$\phi : M \lra R$ we also define two smooth functions on $M$
\ba  %2.3 - 2.4
&&{U}[g,\phi] \equiv g^{MN} (\p_{M} \phi) \p_{N} \phi, \\
&&{V}[g,\phi] \equiv g^{M_{1}N_{1}} g^{M_{2}N_{2}}
[\btd_{M_1}(\p_{M_2} \phi)  +  (\p_{M_1} \phi) \p_{M_2} \phi] \times
\nonumber \\
&&[\btd_{N_1}(\p_{N_2} \phi)  +  (\p_{N_1} \phi) \p_{N_2} \phi],
\ea
where $\btd = {\btd}[g]$ is covariant derivative with respect to $g$.
The scalar invariants (2.1)-(2.3) play an important role in what follows.
Now, we consider the manifold
\be %2.4
M = M_{0} \times M_{1} \times \ldots \times M_{n}
\ee
with the metric
\be %2.5
g= g^{(0)} + \sum_{i=1}^{n} \exp[2{\phi^{i}}(x)] g^{(i)},
\ee
where
$g^{(0)} = {g^{(0)}_{\mu \nu}}(x) dx^{\mu} \otimes dx^{\nu}$ is metric on
$M_{0}$, $g^{(i)}$ is metric on $M_{i}$
and $\phi^{i} : M_0 \lra {\bf R}$ is smooth function, $i = 1, \ldots, n$.

{\bf Proposition 1}. The Riemann tensor squared  for the metric
(2.5) has the following form
\ba %2.6
&&{I}[g] = {I}[g^{(0)}] + \sum_{i=1}^{n} \{ e^{-4 \phi^{i}} {I}[g^{i}]
- 4 e^{-2 \phi^i} {U}[g^{(0)}, \phi^i] {R}[g^{(i)}]
\nonumber \\
&&- 2 N_i {U^2}[g^{(0)}, \phi^i]  + 4 N_i {V}[g^{(0)}, \phi^i] \}
\nonumber \\
&&\sum_{i,j =1}^{n} 2 N_i N_j [g^{(0), \mu \nu} (\p_{\mu} \phi^i)
\p_{\nu} \phi^j]^2 ,
\ea
where  $U$-  and $V$-invariants are defined in
(2.2), (2.3) and ${R}[g^{(i)}]$ is  scalar curvature
of $g^{(i)}$  and  $N_i = {\rm dim} M_i$  is
dimension of $M_i$, $i = 1, \ldots, n$.

{\bf Sketch of proof}. For  $n = 1$ the relation  (2.6)
may be verified by a straightforward calculation. For  $n > 1$
relation (2.6) may  proved by induction (on $n$) using the following
decomposition formulas
\ba %2.7 - 2.8
&&{U}[g^{(0)} + \exp(2{\phi^{1}}(x)) g^{(1)}, {\phi}(x)] =
{U}[g^{(0)}, {\phi}(x)],     \\
&&{V}[g^{(0)} + \exp(2{\phi^{1}}(x)) g^{(1)}, {\phi}(x)] =
{V}[g^{(0)}, {\phi}(x)] +
N_1 [g^{(0), \mu \nu} (\p_{\mu} \phi^1) \p_{\nu} \phi]^2
\ea
($x \in M_0$).

For the scalar curvature of the metric  (2.5) we get
\ba
&&{R}[g] = {R}[g^{(0)}] + \sum_{i =1}^{n} e^{-2 \phi^i} {R}[g^{(i)}] -
\nonumber \\
&&\sum_{i,j =1}^{n} (N_i \delta_{ij} +  N_i N_j)
g^{(0),\mu \nu} (\p_{\mu} \phi^i) \p_{\nu} \phi^j
\nonumber \\
&&- 2 \sum_{i =1}^{n} N_i {\btu}[g^{(0)}] \phi^i,
\ea
where ${\btu}[g^{(0)}]$  is Laplace-Beltrami operator corresponding to
$g^{(0)}$ (see also \cite{Ber}).

Remark 1. In (2.6) and in what follows we use the following
condensed notations: ${I}[g] = {{I}[g]}(x)$,
${I}[g^{(\nu)}] = {{I}[g^{(\nu)}]}(x_{\nu})$,
for $x \in M$, $x_{\nu} \in M_{\nu}$, $\nu = 0, \ldots, n$
and analogously for scalar curvatures.

\section{Multidimensional cosmology}

Here we are interested in the special case of (2.4), (2.5)  with
$M_0 = (t_1, t_2)$, $t_1 < t_2$.
Thus, we consider the metric
\be %3.1
g_c = - {B}(t) dt \otimes dt +  \sum_{i=1}^{n} {A_{i}}(t) g^{(i)},
\ee
defined on the manifold
\be %3.2
M = (t_1, t_2) \times M_{1} \times \ldots \times M_{n}.
\ee
Here, like in (2.4), (2.5) $g^{(i)}$ is a metric on  $M_{i}$ and
${B}(t), {A_{i}}(t) \neq 0$, $i = 1, \ldots, n$.

 From Proposition 1 we obtain the Riemann tensor squared for the metric
(3.1) (see also \cite{IM4})
\ba %3.3
{I}[g_c] =
&& \sum_{i=1}^{n} \{ A_{i}^{-2} {I}[g^{(i)}] + A_{i}^{-3} B^{-1}
\dot{A}_{i}^{2} {R}[g^{(i)}]  - \frac{1}{8}N_{i} B^{-2} A_{i}^{-4}
\dot{A}_{i}^{4}
 \nonumber \\
&&+\frac{1}{4} N_{i} B^{-2}(2 A_{i}^{-1}
\ddot{A}_{i} - B^{-1} \dot{B} A_{i}^{-1} \dot{A}_{i} - A_{i}^{-2}
 \dot{A}_{i}^{2})^{2} \}
 \nonumber \\
&&+ \frac{1}{8}  B^{-2} [\sum_{i=1}^{n} N_{i}
(A_{i}^{-1} \dot{A}_{i})^{2}]^{2}.
\ea
(We recall that ${\rm dim}M_{i} = N_{i}$, $i =1, \ldots , n$.)

For scalar curvature of (3.1) we get from (2.9)
\ba  %3.4
&&{R}[g_c] = \sum_{i=1}^{n} \{ e^{- 2 x^i} {R}[g^{(i)}] +
\nonumber \\
&& e^{- 2 \gamma} N_{i}[ 2 \ddot{x}^i  +
\dot{x}^i (\sum_{j =1}^{n} N_j \dot{x}^j - 2 \dot{\gamma})
+ (\dot{x}^i)^2 ] \},
\ea
where $B = e^{2 \gamma}$  and $A_i = e^{ 2 x^i}$, $i =1, \ldots , n$.

\subsection{$(n+1)$-dimensional Kasner solution}

Let us  consider the metric on ${\bf R}_{+} \times {\bf R}^n$
\be %3.5
g = - dt \otimes dt +  \sum_{i=1}^{n} t^{2 \alpha_i}
 dx^i \otimes dx^i,
\ee
where $t > 0$, $- \infty < x^i < \infty$ and $\alpha_i$ are
constants, $i = 1, \ldots, n$.  From (3.3) we get

\be %3.6
{I}[g] = 2 {F}(\alpha) t^{-4},
\ee where
\be %3.7
{F}(\alpha) =   \sum_{i=1}^{n} [2
\alpha_i^2 (\alpha_i - 1)^2 - \alpha_i^4]  + [\sum_{i=1}^{n}
\alpha_i^2]^2.
\ee
Now we impose the following restrictions on the
parameters  $\alpha_i$
\be %3.8
\sum_{i=1}^{n} \alpha_i =  \sum_{i=1}^{n} \alpha_{i}^2   = 1.
\ee
The metric (3.5) with the restrictions (3.8)
imposed satisfies the vacuum Einstein equations (or, equivalently,
${R_{MN}}[g] = 0$). It is a trivial generalization of
the well-known Kasner solution. In this case
\be %3.9
{F}(\alpha) =  {\Phi}(\alpha) = {\Phi_{n}}(\alpha) \equiv
\sum_{i=1}^{n} [ \alpha_i^4  - 4 \alpha_i^3] + 3.
\ee

We define a Milne set as
\be %3.10
{\cal M} = {\cal M}_n = \{(1, 0, \ldots, 0),  \ldots,
(0, \ldots, 0,1) \}  \subset {\cal E},
\ee
where
\be %3.11
{\cal E} =  {\cal E}_n
\equiv \{ \alpha = (\alpha_1, \ldots \alpha_n) \in
{\bf R}^n | \sum_{i=1}^{n} \alpha_i = \sum_{i=1}^{n} \alpha_{i}^2   = 1
\}.
\ee
Notice, that  ${\cal E}$ is $(n -2)$-dimensional  ellipsoid
for $n >2$ (${\cal E} \simeq {\bf S}^{n-2}$).

For $n=1$, ${\cal M} = {\cal E} = \{ (1) \}$ and we are lead to well-known
Milne solution
\be %3.12
g_M = - dt \otimes dt +  t^{2 } dx^1 \otimes dx^1.
\ee
We recall that by the coordinate transformation
$y^0 = t \cosh x^1$, $y^1 = t \sinh x^1$ the metric (3.12) is reduced
to the Minkowsky metric
$\eta = - dy^0 \otimes dy^0 + dy^1 \otimes dy^1 $
in the upper light cone  $y^0 > | y^1|$.

For $\alpha = (\ldots,0,1_i,0,\ldots) \in {\cal M}$ we
get a trivial extension of the Milne metric:
\be %3.13
g_m = - dt \otimes dt +   t^{2 } dx^i \otimes dx^i
 + \sum_{j \neq i}^{n} dx^j \otimes dx^j,
\ee
$i = 1, \ldots, n$; $n > 1$.

{\bf Proposition 2}. Let $\alpha = (\alpha_1, \ldots \alpha_n)
\in {\cal E}$. Then ${\Phi}(\alpha) \geq 0$  and
${\Phi}(\alpha) = 0$ if and only if $\alpha \in  {\cal M}$.

{\bf Proof}. For $n=1, 2$ the proposition is trivial. So, we
consider the case $n >2$.
Let
\be %3.14
\Phi_{|} = \Phi_{|{\cal E}}: {\cal E}  \lra {\bf R}
\ee
be the restriction of the function $\Phi$ (3.9) on
${\cal E}$ (3.11). Since ${\cal E}$ is a smooth submanifold in ${\bf R}^n$
(see (3.8)) the function $\Phi_{|}$ is also smooth
($\Phi_{|} = \Phi \circ i$, where $i: {\cal E}  \lra {\bf R}^n$ is
canonical embedding). The manifold  ${\cal E}$  is compact (it is
isomorphic to ${\bf S}^{n-2}$). Let ${\rm Min} = {\rm Min}(\Phi_{|})$ is
the set of points of (absolute) minimum of $\Phi_{|}$ and ${\rm Ext} =
{\rm Ext}(\Phi_{|})$ is the set of points of extremum of $\Phi_{|}$.
The set ${\rm Min}$ is
non-empty: ${\rm Min} \neq \emptyset$, since $\Phi_{|}$ is a continuous
real-valued function defined on the compact topological space ${\cal E}$.
Clearly, that ${\rm Min} \subset {\rm Ext}$.

First we find ${\rm Ext}$ using the standard scheme of
conditional extremum.
We consider the function
\be %3.15
{\hat{\Phi}}(\alpha, \lambda, \mu) = {\Phi}(\alpha) +
\mu (\sum_{i=1}^{n} \alpha_{i}^2  - 1) +
\lambda (\sum_{i=1}^{n} \alpha_{i}  - 1)
\ee
where  $\mu, \lambda \in {\bf R} $. The point $\alpha$ belongs to ${\rm
Ext}$ if and only if there exist $\lambda, \mu \in {\bf R}$ such that
$(\alpha, \mu, \lambda)$ is a point of extremum of the function (3.15),
i.e.  the relations (3.8) and
\be %3.16
\frac{\p \hat{\Phi}}{\p \alpha_i}  =
4 \alpha_i^3 - 12 \alpha_i^2  + 2 \mu \alpha_i + \lambda = 0,
\ee
$i = 1, \ldots, n$, are satisfied.
 From (3.8) and (3.16)  we obtain
\ba %3.17 -18
&&4 \sum_{i=1}^{n} \alpha_i^3 - 12 + 2 \mu  + \lambda n = 0, \\
&&4 \sum_{i=1}^{n} \alpha_i^4 - 12 \sum_{i=1}^{n} \alpha_i^3
+ 2 \mu  + \lambda  = 0,
\ea
and hence
\be %3.19
4{\Phi}(\alpha) =   \lambda (n-1).
\ee
Let us consider the cubic equation
\be %3.20
4 y^3 - 12 y^2  + 2 \mu y + \lambda = 4 (y -y_1)(y -y_2)(y -y_3)=  0.
\ee
We prove that for given $\mu$ and $\lambda$ the cubic equation (3.20)
should have three different real roots. Indeed, there should be at least
two different real roots since otherwise $\alpha_1 = \ldots = \alpha_n$
but this is impossible due to the Kasner constraints (3.8). The third
root should be also real and we are lead to the following
three possibilities for the roots: i)$ y_1 = y_2 < y_3$;
ii)$y_1 < y_2 = y_3$;  iii)$ y_1 < y_2 < y_3$.
It follows from (3.20) that
\be %3.21
y_1 + y_2 + y_3 = 3
\ee
and hence  $y_3 > 1$. But due to (3.8) $\alpha_i \leq 1$ and as a
consequence the possibilities i) and ii) can not be fulfilled
(otherwise $\alpha_i = y_1$ for all $i$). Thus  $y_1 < y_2 < y_3$  and due
to (3.16) and $y_3 > 1$ we obtain
\be %3.22
\{x| x = \alpha_i, i =1, \ldots , n  \} =    \{ y_1, y_2 \}.
\ee
Solving  (3.8) for $\alpha$ satisfying (3.22) we get
\ba %3.23 -25
&&y_1 = {y_1}(m_1,m_2) = \frac{m_1 - \sqrt{\Delta}}{m_1 n}, \\
&&y_2 = {y_2}(m_1,m_2)= \frac{m_2 + \sqrt{\Delta}}{m_2 n}, \\
&& \Delta = m_1 m_2 (n - 1).
\ea
Here $m_a$ is the number of $\alpha_i$ equal to $y_a$, $a = 1,2$.
Clearly, that $m_1 + m_2 = n$ and $m_1, m_2 \geq 1$.
It follows from (3.19) and (3.20) that
\ba %3.26 -27
&&\lambda = 4{\Phi}(\alpha)/(n-1)  = - 4  y_1 y_2 y_3, \\
&&\mu = 2 (y_1 y_2 + y_2 y_3 + y_3 y_1).
\ea
In fact we find the expression for  the  set of extremum
\be %3.28
{\rm Ext} = E_1 \sqcup \ldots \sqcup E_n,
\ee
where
\be %3.29
E_k = \{ (\alpha_1= y_2, \ldots, \alpha_k= y_2,  \alpha_{k + 1}= y_1,
\ldots, \alpha_n= y_1) \  \rm{and \ all \ permutations} \},
\ee
$k = 1, \ldots, n-1$  and $y_1 = {y_1}(n-k,k)$, $y_2 = {y_2}(n-k,k)$
(see (3.23), (3.24)).  Clearly, that the number of
elements in ${\rm Ext}$ is $2^n -2$.

For $\alpha \in E_k$, $k >1$, we have
\be %3.30
{\Phi}(\alpha) > 0.
\ee
This can be readily verified using the inequalities
${y_1}(n-k,k) < 0$, $0 < {y_2}(n-k,k) < {y_3}(n-k,k)$, $k >1$,
and the relation (3.26). For $\alpha \in E_1$
\be %3.31
{\Phi}(\alpha) = 0.
\ee
Using the inclusion ${\rm Min} \subset {\rm Ext}$  and the relations
(3.28), (3.30) and (3.31) we obtain
\be %3.32
{\rm Min} = {\rm Min}(\Phi_{|}) = E_1 =
{\cal M}.  \ee The proposition 2 follows from the relations (3.31) and
(3.32).

Remark 2. It will be proved in a separate publication
that  $\Phi_{|}$ (see (3.14)) is a Morse function.

 From (3.6), (3.9) and Proposition 2
we get for the Kasner metric  (3.5) with  $\alpha \in
{\cal E} \setminus {\cal M}$
\be %3.33
{{I}[g]}(t,{\vec{x}}(t))  \ra + \infty,
\qquad {\rm as} \  t \ra +0.
\ee
for arbitrary curve ${\vec{x}}(t)$.

\subsection{Kasner-like solutions with Ricci-flat spaces}

Here we consider the following metric \cite{I}
\be % 3.34
g= - dt \otimes dt +
\sum_{i=1}^{n} t^{2 \alpha_i} c_i g^{(i)},
\ee
defined on the manifold (3.2) with $(t_1, t_2) = (0, +\infty) =
{\bf R}_{+}$ ,
where $(g^{(i)}, M_i)$ are Ricci-flat internal spaces,
i.e. ${R_{m_i n_i}}[g^{(i)}] =0$, and  $c_i \neq 0$ are constants,
$i = 1, \ldots, n$, $n \geq 2$.
Parameters $\alpha_i$ satisfy the relations
\be %3.35
\sum_{i=1}^{n} N_i \alpha_i =  \sum_{i=1}^{n} N_i \alpha_{i}^2   = 1.
\ee
The metric (3.34) with the restrictions (3.35)
imposed satisfies the vacuum Einstein equations.

For the metric  (3.34), (3.35) we get from (3.3)
\be % 3.36
{I}[g] = \sum_{i=1}^{n}  t^{-4 \alpha_i} c_i^{-2} {I}[g^{(i)}] +
2 {\Phi_{*}}(\alpha) t^{-4}, \ee
where
\be %3.37
{\Phi_{*}}(\alpha) \equiv
\sum_{i=1}^{n} N_i [ \alpha_i^4  - 4 \alpha_i^3] + 3.
\ee
Analogously to (3.10) we introduce the
Milne set
\be %3.38
{\cal M_{*}} = \{ \alpha| \alpha =
(\ldots, 0, 1_i, 0, \ldots), N_i = 1 \} \subset {\cal E}_{*},
\ee
where
\be %3.39
{\cal E}_{*} \equiv \{ \alpha = (\alpha_1, \ldots \alpha_n) \in
{\bf R}^n | \sum_{i=1}^{n} N_i \alpha_i = \sum_{i=1}^{n} N_i \alpha_{i}^2
= 1 \}.
\ee
For $n> 2$ ${\cal E_{*}}$ is $(n-2)$-dimensional ellipsoid.

Example 1. The set (3.38) is empty: ${\cal M_{*}} = \emptyset$,
if and only if $N_i > 1$ for all $i$.

Example 2. For  $N_1 = \ldots = N_n = 1$ we have ${\cal M_{*}} = {\cal M}$
(see (3.10)).

{\bf Proposition 3}. Let $\alpha = (\alpha_1, \ldots \alpha_n)
\in {\cal E_{*}}$. Then ${\Phi_{*}}(\alpha) \geq 0$  and
${\Phi_{*}}(\alpha) = 0$ if and only if $\alpha \in  {\cal M_{*}}$.

{\bf Proof}. Here we consider the function  $\Phi = \Phi_{N}$ (3.9)
corresponding to  $N = \sum_{i=1}^{n} N_i$.  For
$\alpha \in  {\cal E_{*}}$ we have
\be %3.40
{\Phi_{*}}(\alpha) = {\Phi_{N}}({\beta}(\alpha)),
\ee
where the set ${\beta}(\alpha) = \beta = (\beta_1, \ldots, \beta_N)$ is
defined by the relations
\ba %3.41
&&\beta_1= \ldots = \beta_{N_1} = \alpha_1,  \nonumber \\
&&\ldots  \nonumber \\
&&\beta_{N - N_n +1}= \ldots = \beta_{N} = \alpha_n.
\ea
It is evident that $\beta \in  {\cal E}_N$ (see (3.11)).
The Proposition 3 follows from the Proposition 2, relation (3.40)
and the equivalence
\be %3.42
{\beta}(\alpha) \in {\cal M}
\Longleftrightarrow
\alpha \in {\cal M}_{*}.
\ee
Here ${\cal M} = {\cal M}_{N}$ is the Milne set corresponding to
$N$.

{\bf Proposition 4}. Let $g$ be the metric (3.34) with the set
$\alpha = (\alpha_1, \ldots \alpha_n) \in {\cal E_{*}}
\setminus {\cal M_{*}}$ and
\be % 3.43
{I}[g^{(i)}] \geq 0,
\ee
for all  $i = 1, \ldots, n$. Then
\be %3.44
{{I}[g]}(t, {f}(t)) \ra + \infty, \qquad {\rm as} \ t \ra +0 ,
\ee
for any function
\be %3.45
f: {\bf R}_{+}  \lra M_{1} \times \ldots \times M_{n}.
\ee
If the condition (3.43) is not imposed the relation (3.44)
takes place if ${f}(t)  \ra f_0 \in M_{1} \times \ldots \times M_{n}$
as $ t \ra +0 $.

{\bf Proof}. The first part of the proposition follows from (3.36), (3.43)
and the inequality  ${\Phi_{*}}(\alpha) > 0$ for
$\alpha \notin {\cal M_{*}}$ (see proposition 3).
The second part of the proposition follows from continuity of
the functions  ${I}[g^{(i)}]$ on $M_i$, the inequalities $\alpha_i < 1$,
 $i = 1, \ldots, n$, and the relation (3.36). Indeed,
the functions ${{I}[g^{(i)}]}({f^i}(t))$, $i = 1, \ldots, n$,
have limits as $t \ra +0 $, and hence are bounded. Here
${f}(t)= ({f^1}(t), \ldots , {f^n}(t)$). Thus, the second term in
the right hand side of (3.36) is dominating in the limit $t \ra +0 $
and we are lead to (3.44).

\subsection{The solutions with asymptotically Kasner behaviour}

Here we consider the metric
\be %3.46
g= - w d \tau \otimes d \tau
+ \sum_{i=1}^{n} {A_{i}}(\tau) g^{(i)},
\ee
defined on the manifold
\be %3.47
(0, T) \times M_{1} \times \ldots \times M_{n},
\ee
where $T >0$ and
$w = \pm 1$. We suppose that the metric $g^{(i)}$ on the
manifold $M_i$ satisfy the following conditions:
the functions ${I}[g^{(i)}]$, $w {{R}[g^{(i)}]}$,
are bounded from below, i.e
\be  %3.48
{{I}[g^{(i)}]}(x_i) \geq C_i
\ee
for all $x_i \in M_i$ and
\be  %3.49
w {{R}[g^{(i)}]}(x_i) \geq D_i
\ee
for all $x_i \in M_i$, $i = 1, \ldots, n$.

Remark 3. Clearly that the conditions (3.48), (3.49) are
satisfied for compact manifold $M_i$. The first condition
is also satisfied when the metric $g^{(i)}$ has the Euclidean
signature: in this case $C_i =0$.

We also suppose that the scale factors $A_{i}: (0, T) \lra {\bf R}$
are smooth functions (${A_{i}}(\tau) \neq 0$),
satisfying the following asymptotical relations
\ba %3.50-52
&&{A_{i}}(\tau)  =  c_i \tau^{2 \alpha_i}[1 + {o}(1)], \\
&&{\dot{A}_{i}}(\tau)  =  c_i \tau^{2 \alpha_i - 1}
[2 \alpha_i + {o}(1)], \\
&&{\ddot{A}_{i}}(\tau)  =  c_i \tau^{2 \alpha_i - 2}
[2 \alpha_i (2 \alpha_i -1) + {o}(1)],
\ea
as $\tau \ra +0$, where $c_i \neq 0, \alpha_i$ are constants,
$i = 1, \ldots, n$. We recall that the notation
${\varphi}(\tau) = {o}(1)$ as $\tau \ra +0$
means that
${\varphi}(\tau) \ra 0$ as $\tau \ra +0$.

Remark 4. The relations (3.51) and (3.52) should not obviously
follow from (3.50). A simple counterexample is
\be %3.53
{A_{i}}(\tau)  = 1 +  \tau \sin \frac{1}{\tau^2}.
\ee
But if
\be %3.54
{A_{i}}(\tau)  =  c_i \tau^{2 \alpha_i}[1 + {\varphi_i}(\tau)],
\ee
where  $c_i \neq 0$ and
\be %3.55
{\varphi_i}(\tau) = {o}(1), \qquad
\tau {\dot{\varphi_i}}(\tau) = {o}(1), \qquad
\tau^2 {\ddot{\varphi_i}}(\tau) = {o}(1),
\ee
as $\tau \ra +0$, $i = 1, \ldots, n$, then the relations (3.50)-(3.52) are
satisfied.

{\bf Theorem}. Let $g$ be the metric (3.46) defined on the manifold
(3.47), where the metrics $g^{(i)}$, $i = 1, \ldots, n$, satisfy
the relations (3.48) and (3.49). Let the scale factors  ${A_{i}}(\tau)$,
$i = 1, \ldots, n$, satisfy the relations (3.50)-(3.52),
where the set of parameters $\alpha = (\alpha_i)$
satisfies the Kasner-like relations  (3.35) ($N_i = {\rm dim} M_i$)
and is non-exceptional, i.e. $\alpha \notin {\cal M_{*}}$
(${\cal M_{*}}$ is defined in (3.38)).
Then
\be %3.56
{{I}[g]}(\tau, x) \ra + \infty, \qquad {\rm as} \ \tau \ra +0,
\ee
uniformly on $x \in  M_{1} \times \ldots \times M_{n}$.

{\bf Proof}.  From (3.3) we get
\be %3.57
{I}[g] =  {I_1}[g] + {I_2}[g] +  {I_3}[g],
\ee
where
\ba %3.58-60
&&{I_1}[g] = \sum_{i=1}^{n} A_{i}^{-2} {I}[g^{(i)}], \\
&&{I_2}[g] =  \sum_{i=1}^{n} A_{i}^{-3}  \dot{A}_{i}^{2} w {R}[g^{(i)}], \\
&&{I_3}[g] = \sum_{i=1}^{n} \{ - \frac{1}{8} N_{i} A_{i}^{-4}
\dot{A}_{i}^{4} + \frac{1}{4} N_{i} (2 A_{i}^{-1} \ddot{A}_{i}
- A_{i}^{-2} \dot{A}_{i}^{2})^{2} \} \nonumber \\
&&+ \frac{1}{8} [\sum_{i=1}^{n} N_{i} (A_{i}^{-1} \dot{A}_{i})^{2}]^{2}.
\ea
 From (3.50)-(3.52) and (3.60) we obtain
\be % 3.61
{I_3}[g] =  [2 {\Phi_{*}}(\alpha) + {o}(1)] \tau^{-4},
\ee
as $\tau \ra +0$, where ${\Phi_{*}}(\alpha)$ is defined in (3.37).
We note that due to $\alpha \notin {\cal M_{*}}$ and Proposition 3
\be %3.62
{\Phi_{*}}(\alpha) >0.
\ee
 From  (3.50), (3.51) we get
\be %3.63
\frac{\dot{A}_i^2}{A_i^3}  =  c_i^{-1} \tau^{-2 - 2 \alpha_i}
[4 \alpha_i^2 + {o}(1)],
\ee
as $\tau \ra +0$.
We note also that due to  $\alpha \notin {\cal M_{*}}$
\be %3.64
\alpha_i < 1,
\ee
$i = 1, \ldots, n$.
Let
\be %3.65
\delta = \min_{i} (1 - \alpha_i) > 0.
\ee
Then it follows from (3.63) that  there exists $\tau_1 >0$
such that
\be %3.66
0 \leq \frac{\dot{A}_i^2}{A_i^3}  < \tau^{- 4 + \delta},
\ee
for all $\tau < \tau_1$, $i = 1, \ldots, n$.
 From (3.49) and (3.66) we get
\be  %3.67
\frac{\dot{A}_i^2}{A_i^3}
w {{R}[g^{(i)}]}(x_i) \geq -|D_i| \tau^{- 4 + \delta}
\ee
for all $x_i \in M_i$, $\tau < \tau_1$, $i = 1, \ldots, n$, and
hence
\be %3.68
{{I_2}[g]}(\tau,x)  \geq -A \tau^{- 4 + \delta},
\ee
for all  $\tau < \tau_1$  and $x \in  M_{1} \times \ldots \times M_{n}$
($A = \sum_{i=1}^{n} |D_i|$). Analogously we get from (3.50)
\be  %3.69
A_i^{-2}  =  c_i^{-2} \tau^{- 4 \alpha_i}
[1 + {o}(1)],
\ee
and consequently there exists $\tau_2 >0$ such that
\be  %3.70
A_i^{-2}  < \tau^{- 4 + \delta}
\ee
for all $\tau < \tau_2$, $i = 1, \ldots, n$.
Using (3.48) and (3.70) we obtain (analogously to (3.68))
\be  %3.71
{{I_1}[g]}(\tau,x)  \geq - B \tau^{- 4 + \delta},
\ee
for all  $\tau < \tau_2$  and $x \in  M_{1} \times \ldots \times M_{n}$
($B = \sum_{i=1}^{n} |C_i|$).
It follows from (3.61),  (3.68), (3.71) that
\be  %3.72
{{I}[g]}(\tau,x)  \geq  \tau^{- 4} [2 {\Phi_{*}}(\alpha)
- (A + B) \tau^{\delta} + {o}(1)] \ra + \infty,
\ee
as  $\tau \ra +0$. This imply the relation (3.56). Theorem is proved.

Remark 5. When the relations (3.48) and (3.49) are not imposed the
relation (3.56) is valid (at least) for any (fixed)
$x \in  M_{1} \times \ldots \times M_{n}$.

\section{Spherically symmetric solutions with Ricci-flat internal spaces}

Now we apply the obtained above results to
the following scalar vacuum solution \cite{IM3}
\ba %4.1-4.2
g= &&- f^{a} dt \otimes dt + f^{b-1}  dR \otimes dR
+ f^{b} R^{2} d \Omega^{2}_{d} +  \sum_{i=1}^{n} f^{a_{i}}
B_i g^{(i)}, \\
&&\exp(2 \varphi)= B_{\varphi}  f^{a_{\varphi}},
\ea
defined on the manifold
\be %4.3
M = (R_0, + \infty) \times {\bf R} \times {\bf S}^{d}  \times
M_{1} \times \ldots \times M_{n},
\ee
where $(M_{i}, g^{(i)})$ are Ricci-flat internal spaces,
${\rm dim} M_{i} = N_{i}$, $i = 1, \ldots, n$,
$d \Omega^{2}_{d}$
is the canonical metric on $d$-dimensional sphere
${\bf S}^{d}$ ($d \geq 2$) and  $f = {f}(R) = 1 - (R_0/ R)^{d-1}$.
Here $R_0, B_{\varphi}, B_i > 0$ are  constants
and the parameters $b, a, a_{1}, \ldots , a_{n}$ satisfy
the relations
\ba  %4.4 -4.5
&&b = (1 - a - \sum_{i=1}^{n} a_{i}N_{i})/(d-1),  \\
&&(a + \sum_{i=1}^{n} a_{i}N_{i})^{2} +
(d-1) (a^{2} + a_{\varphi}^{2} +
\sum_{i=1}^{n} a_{i}^{2} N_{i})= d.
\ea
The solution (4.1)-(4.3) is a scalar-vacuum multispace
generalization  of the Tangherlini solution \cite{T}.
In the parametrization of the harmonic-type variable this solution was
presented earlier in \cite{BM,IM4}.
For $a_{\varphi} = 0$ see also \cite{FIM,IM4}. Some special cases were
considered earlier in \cite{BIM} (for $d=2$, $a_{\varphi} = 0$) and
\cite{BI} ($n = 1$ and $d=2$).

The metric and scalar field from (4.1), (4.2) satisfy the field
equations
\ba %4.6-4.7
&&{R_{MN}}[g] = \p_{M} \varphi \p_{N} \varphi, \\
&& {\btu}[g] \varphi = 0,
\ea
corresponding to the action
\be   %4.8
S = \int d^{D}x \sqrt{|g|} \{
{R}[g] -  \partial_{M} \varphi
\partial_{N} \varphi g^{MN} \}.
\ee

Now, we introduce a new variable
\be   %4.9
\tau = {\tau}(R) = \int_{R_0}^{R}  dx [{f}(x)]^{(b-1)/2}.
\ee
The integral in (4.9) is convergent since due (4.4) and (4.5)
\be   %4.10
b > -1.
\ee
The map (4.9) defines a diffeomorphism from $(R_0, + \infty)$ to
${\bf R}_{+}$. We consider the diffeomorphism
\be   %4.11
\sigma : M^{'} \lra M,
\ee
generated by (4.9): ${\sigma}(\tau,t, \ldots) = ({R}(\tau),t, \ldots)$,
where
\be %4.12
M^{'} = {\bf R}_{+} \times {\bf R} \times {\bf S}^{d}  \times M_{1} \times
\ldots \times M_{n}.
\ee
The substitution (4.9) into (4.1) gives a metric on the manifold
(4.12) (the dragging of (4.1) by the map (4.11))
\ba %4.13
\sigma^{*}g=  d \tau \otimes d \tau
+ {A_{0}}(\tau) g^{(0)} \nonumber \\
+ \sum_{i=1}^{n} {A_{i}}(\tau)
g^{(i)} - {A_{-1}}(\tau) dt \otimes dt,
\ea
where $g^{(0)} = d \Omega^{2}_{d}$ and
\ba %4.14 -15
&&{A_{i}}(\tau)  = [{f}({R}(\tau))]^{a_i}, \\
&&{A_{0}}(\tau)  = {R^2}(\tau)[{f}({R}(\tau))]^{b},
\ea
$i = -1, 1, \ldots, n$; $a_{-1} = a$.
 From (4.9) and the asymptotical behaviour
\be %4.16
{f}(R) \sim \frac{(d-1)}{R_0} (R- R_0),  \qquad {\rm as} \ R \ra R_0
\ee
we get
\ba %4.17
R- R_0  \sim (c_{*} \tau)^{2/(b+1)},   \qquad
{f}({R}(\tau)) \sim c_{.} \tau^{2/(b+1)},
\ea
as $\tau \ra +0$, where $c_{*}, c_{.}$ are constants.
 From (4.17) we get for the scale factors (4.14), (4.15)
and the scalar field the following asymptotical relations
\ba %4.18 -20
&&{A_{i}}(\tau)  \sim  c_i \tau^{2 \alpha_i}, \\
&&{A_{0}}(\tau)  \sim  c_0 R^2_0 \tau^{2 \alpha_0}, \\
&&\exp(2{\varphi}(\tau))  \sim  c_{\varphi} \tau^{2 \alpha_{\varphi}},
\ea
as $\tau \ra +0$, where $c_{i}, c_{0}, c_{\varphi}$ are constants,
and
\ba %4.21 -22
&&\alpha_i  = a_i/(b+1), \qquad \alpha_0 = b/(b+1), \\
&&\alpha_{\varphi}  = a_{\varphi}/(b+1),
\ea
$i = -1, 1, \ldots, n$. The parameters  (4.21), (4.22)
are correctly defined due to (4.10) and satisfy the
Kasner-like relations
\ba %4.23 -24
&&\sum_{\nu=-1}^{n} N_{\nu} \alpha_{\nu} = 1, \\
&&\sum_{\nu =-1}^{n} N_{\nu} \alpha_{\nu}^2 + \alpha_{\varphi}^2= 1.
\ea
Here $N_{-1} =1$ and  $N_0 =d$.

Now we consider the case $\alpha_{\varphi} = 0$ (or equivalently
$a_{\varphi} = 0$). Let
\ba %4.25 26
&&{\cal M}_{1} = \{ \alpha = (\alpha_{-1}, \ldots, \alpha_n) =
(\ldots, 0, 1_{\nu}, 0, \ldots), N_{\nu} = 1 \}
\subset {\bf R}^{n+2}, \\
&&{\cal T} = \{ a_{.} = (a_{-1}, a_1, \ldots, a_n) =
(\ldots, 0, 1_{\nu}, 0, \ldots), N_{\nu} = 1 \}
\subset {\bf R}^{n+1}.
\ea
Clearly, that  ${\cal M}_1 \subset {\cal E}_2$ and
${\cal T} \subset {\cal E}_1$, where ${\cal E}_1 \subset {\bf R}^{n+1}$
and ${\cal E}_2 \subset {\bf R}^{n+2}$ are $n$-dimensional ellipsoids
defined by relations (4.5) and (4.23), (4.24) respectively. It is not
difficult to verify that the function  $\alpha = {\alpha}(a_{.})$
from (4.21) defines the diffeomorphism  ${\cal E}_1  \ra {\cal E}_2$ and
\be %4.27
{\alpha}(a_{.})  \in {\cal M}_{1}  \Longleftrightarrow   a_{.}
\in {\cal T}.  \ee

{\bf Proposition 5}. Let $\sigma^{*} g$ be the metric (4.13)-(4.15) with
the parameters (4.21), (4.22) satisfying the relations (4.23), (4.24)
and obeying the restrictions: $\alpha_{\varphi} = 0$,
$\alpha = (\alpha_{-1}, \ldots \alpha_n) \notin {\cal M}_1$.
Let Ricci-flat internal spaces $(M_{i}, g^{(i)})$, $i = 1, \ldots, n$,
satisfy the self-boundness conditions (3.48).  Then
\be %4.28
{{I}[\sigma^{*}g]}(\tau, y) \ra + \infty, \ {\rm as} \  \tau \ra +0,
\ee
uniformly on $y \in {\bf R} \times {\bf S}^d \times M_{1} \times \ldots
\times M_{n}$.

{\bf Proof}. We denote $(M_{-1}, g^{(-1)}) = ({\bf R}, - dt \otimes dt)$
and  $(M_{0}, g^{(0)}) = ({\bf S}^d, d \Omega^{2}_{d})$.
Due to assumption of the proposition, flatness of $(-1)$-space
and the relations
\be %4.29
{I}[g^{(0)}]   = 2d(d-1) = 2 {R}[g^{(0)}],
\ee
the conditions of the Theorem ((3.48), (3.49)) are satisfied for
all spaces $(M_{\nu}, g^{(\nu)})$, $\nu = -1, \ldots, n$.
All scale factors ${A_{\nu}}(\tau)$, $\nu = -1, \ldots, n$,
and their first and second derivatives satisfy the Kasner-like
asymptotical conditions of the theorem (see (3.50)-(3.52)).
This may be proved using asymptotical relations (4.18), (4.19)
and (4.9). Thus, the proposition 5 follows from the Theorem.

Using equivalence (4.27) and the relation
${{I}[\sigma^{*}g]}(\tau, y) = {{I}[g]}({R}(\tau), y)$ we may
reformulate the Proposition 5 for the metric (4.1).

{\bf Proposition 6}. Let $g$ be the metric (4.1) with the
parameters  satisfying (4.4), (4.5) and $a_{\varphi} = 0$,
$(a , a_{1}, \ldots a_n) \notin {\cal T}$ (see (4.26)).
Let Ricci-flat internal spaces $(M_{i}, g^{(i)})$, $i = 1, \ldots, n$,
satisfy the self-boundness conditions (3.48).  Then
\be %4.30
{{I}[g]}(R, y) \ra + \infty, \ {\rm as} \ R  \ra R_0,
\ee
uniformly on $y \in {\bf R} \times {\bf S}^d \times M_{1} \times \ldots
\times M_{n}$.

Remark 6. Due to (3.3), (4.29)  and the flatness of $t$-space
{I}[g] does not depend on $y_j \in M_j$, $j = -1, 0$.

Remark 7. From (4.6) we obtain
\be %4.31
{R}[g] = g^{MN} \p_{M} \varphi \p_{N} \varphi
= a_{\varphi}^2  f^{ -1 -b} (f')^2.
\ee
where $f = {f}(r) = 1 - (R_0/ r)^{d-1}$.
Using (4.31), (4.10) and the relation
\be %4.32
f' = (d-1) R_0^{d-1} r^{-d}
\ee
we obtain  for $a_{\varphi} \neq 0$
\be  %4.33
{{R}[g]}(r, y) \ra + \infty, \ {\rm as} \  r  \ra R_0.
\ee

\subsection{Multitemporal generalization of Tangherlini solution}

Now we consider the special case of the solution (4.1)-(4.3) with
 $n-1$ one-dimensional internal spaces (extra times).
This solution
defined on the manifold
\be %4.34
M = (R_0, + \infty) \times {\bf R}^n \times {\bf S}^{d},
\ee
reads
\ba %4.35-36
g= &&- \sum_{i=1}^{n} f^{a_{i}}  B_i dt^i \otimes dt^i +
f^{b-1}  dR \otimes dR + f^{b} R^{2} d \Omega^{2}_{d} , \\
&&\exp(2 \varphi)= B_{\varphi}  f^{a_{\varphi}},
\ea
$f = {f}(R) = 1 - (R_0/ R)^{d-1}$, where
$R_0, B_{\varphi}, B_i > 0$ are  constants
and the parameters $b, a_{1}, \ldots , a_{n}$ satisfy
the relations
\ba  %4.37 -4.38
&&b = (1 - \sum_{i=1}^{n} a_{i})/(d-1),  \\
&&(\sum_{i=1}^{n} a_{i})^{2} +
(d-1) (a_{\varphi}^{2} +
\sum_{i=1}^{n} a_{i}^{2})= d.
\ea

Let us consider the case  $a_{\varphi} =0$ \cite{IM4}.
The "Tangherlini set" in thi case (see (4.26))
\be %4.39
{\cal T} = \{(1, 0, \ldots, 0),  \ldots,
(0, \ldots, 0,1) \}  \subset {\bf R}^n.
\ee
consists of $n$ points. As was shown in \cite{IM4} the multitemporal
horizon takes place only for  $(a_1, \ldots, a_n) \in {\cal T}$.
For $(a_1, \ldots, a_n) \notin {\cal T}$ we get from Proposition 6
\be %4.40
{{I}[g]}(R) \ra + \infty, {\rm as} \ R  \ra R_0,
\ee
i.e. the solution (4.35) describes a multitemporal naked singularity.
(The relation (4.40) was stated previously in \cite{IM4}.)
This singular solution is unstable under monopole perturbations:
this follows from the recent (more general) result of Bronnikov {\it
et  al} \cite{B,BM,BBMF}.

When  $a_{\varphi} \neq 0$, we get from  (4.33)
\be  %4.41
{{R}[g]}(r) \ra + \infty, {\rm as} \ r  \ra R_0.
\ee
It may be shown that in this case we also have a
multitemporal naked singularity. It should be noted  also
that for  the case $n =2$ the considered multitemporal
solutions were recently generalized in \cite{B} for more
complicated model.

\begin{center} {\bf Acknowledgement}    \end{center}

One of the authors (V.D.I) is grateful to A.F.Ionov, Yu.S. Vladimirov,
and  organizers of the First Ionov School in Yaroslavl  for
invitation and kind hospitality.

\pagebreak

\end{document}